\documentclass[a4paper,12pt]{article}

\usepackage{ifpdf}
\newif\ifpdf
\ifx\pdfoutput\undefined
  \pdffalse
\else
  \pdfoutput=1
  \pdftrue
\fi
\RequirePackage{xspace} %
\RequirePackage{subfigure} %
\RequirePackage[centertags]{amsmath} %
\RequirePackage{amssymb}
\RequirePackage{wrapfig} %
\RequirePackage{calc} %
\RequirePackage{ifthen}
\RequirePackage{tabularx} %
\RequirePackage{flafter} %
\RequirePackage{fancyhdr} %
\ifpdf
  \RequirePackage[pdftex]{color}%
  \RequirePackage{colortbl}%
  \RequirePackage{array}%
  \RequirePackage[pdftex]{graphicx}
  \RequirePackage[ pdftex, plainpages = false, pdfpagelabels,
                 pdfpagelayout = useoutlines,
                 bookmarks,
                 breaklinks = true,
                 linktocpage,
                 pagebackref,                      
                 colorlinks = true,
                 linkcolor = blue,
                 urlcolor  = blue,
                 citecolor = blue,
                 anchorcolor = blue,
                 hyperindex = true,
                 hyperfigures
                 ]{hyperref}
\else
  \RequirePackage{color}
  \RequirePackage{colortbl}
   \RequirePackage{array}
  \RequirePackage[dvips]{graphicx}
  \RequirePackage{hyperref}
  \usepackage{rotating}
\fi
\usepackage{makeidx}
\usepackage{setspace}
\usepackage{rotating}
\usepackage{ecltree}
\usepackage{epic}
\usepackage{supertabular}
\usepackage{color}
\usepackage{exscale}
\usepackage{fontenc}
\usepackage{ifthen}
\usepackage{latexsym}
\usepackage{makeidx}
\usepackage{syntonly}
\usepackage{inputenc}
\usepackage{graphicx}
\usepackage{setspace}
\usepackage{caption2}
\usepackage[english]{babel}
\usepackage[square, comma,numbers,sort&compress]{natbib}
\usepackage{hypernat}
\usepackage{boxedminipage}
\usepackage{framed}
\usepackage{longtable}
\usepackage[all]{hypcap}

\newcommand{\Hs}      {\hspace{-0.5cm}} %
\newcommand{\CIF}     {\centering \includegraphics[width=2.5in]} %
\newcommand{\Vmin}    {\vspace{-0.2cm}} %

\newcommand{\note}[1]{\marginpar[left]{\singlespace \tiny #1}}
\renewcommand{\sectionmark}[1]%
      {\markright{\thesection\ #1}} 

\renewcommand{\note}[1]{}
\newcommand{\arccosh}     {{\rm arccosh}}
\newcommand{\etal}     {{\it et al.}}

\doublespace

\title
{ %
\vspace*{3.0cm} \LARGE{\bf Newtonian Flow in Converging-Diverging Capillaries} \vspace*{4.0cm} \\
}

\author{Taha Sochi\footnote{Imaging Sciences \& Biomedical Engineering, King's College London, The Rayne
Institute, St Thomas' Hospital, London, SE1 7EH, UK. Email: taha.sochi@kcl.ac.uk.} \vspace*{5.0cm}}

\begin{document}

\maketitle %
\pagenumbering{arabic}

\newpage
\phantomsection \addcontentsline{toc}{section}{Contents} %
\tableofcontents

\newpage
\phantomsection \addcontentsline{toc}{section}{List of Figures} %
\listoffigures

\phantomsection \addcontentsline{toc}{section}{List of Tables} %
\listoftables

\newpage
\phantomsection \addcontentsline{toc}{section}{Abstract} \noindent
{\noindent \LARGE \bf Abstract} \vspace{0.5cm}\\
\noindent %

The one-dimensional Navier-Stokes equations are used to derive analytical expressions for the
relation between pressure and volumetric flow rate in capillaries of five different
converging-diverging axisymmetric geometries for Newtonian fluids. The results are compared to
previously-derived expressions for the same geometries using the lubrication approximation. The
results of the one-dimensional Navier-Stokes are identical to those obtained from the lubrication
approximation within a non-dimensional numerical factor. The derived flow expressions have also
been validated by comparison to numerical solutions obtained from discretization with numerical
integration. Moreover, they have been certified by testing the convergence of solutions as the
converging-diverging geometries approach the limiting straight geometry.


\pagestyle{headings} %
\addtolength{\headheight}{+1.6pt}
\lhead[{Chapter \thechapter \thepage}]%
      {{\bfseries\rightmark}}
\rhead[{\bfseries\leftmark}]%
     {{\bfseries\thepage}} 
\headsep = 1.0cm               

\newpage
\section{Introduction} \label{Introduction}

Modeling the flow through capillaries of converging-diverging geometries is an important subject
and has many scientific and industrial applications. Moreover, it is required for modeling
viscoelasticity, yield-stress and the flow of Newtonian and non-Newtonian fluids through porous
media \cite{SochiB2008, SochiVE2009, SochiFeature2010, SochiYield2010, SochiComp2010,
SochiArticle2010, SochiPower2011, SochiSlip2011}.

There are many previous attempts to model the flow through capillaries of various geometries.
However, they either apply to tubes of regular cross sections \cite{Whitebook1991, SisavathJZ2001}
or deal with very special cases. Most these studies use numerical meshing techniques such as finite
difference and spectral methods to obtain numerical results. Some examples of these attempts are
Kozicki \etal\ \cite{KozickiCT1966}, Miller \cite{Miller1972}, Oka \cite{Oka1973}, Williams and
Javadpour \cite{WilliamsJ1980}, Phan-Thien \etal\ \cite{ThienGB1985, ThienK1987}, Lahbabi and Chang
\cite{LahbabiC1986}, Burdette \etal\ \cite{BurdetteCAB1989}, Pilitsis \etal\ \cite{PilitsisSB1991,
PilitsisB1989}, James \etal\ \cite{JamesTKBP1990}, Talwar and Khomami \cite{TalwarK1992}, Koshiba
\etal\ \cite{KoshibaMNS2000}, Masuleh and Phillips \cite{MasulehP2004}, and Davidson \etal\
\cite{DavidsonLC2008}.

\begin{figure}[!h]
\centering{}
\includegraphics
[scale=0.6] {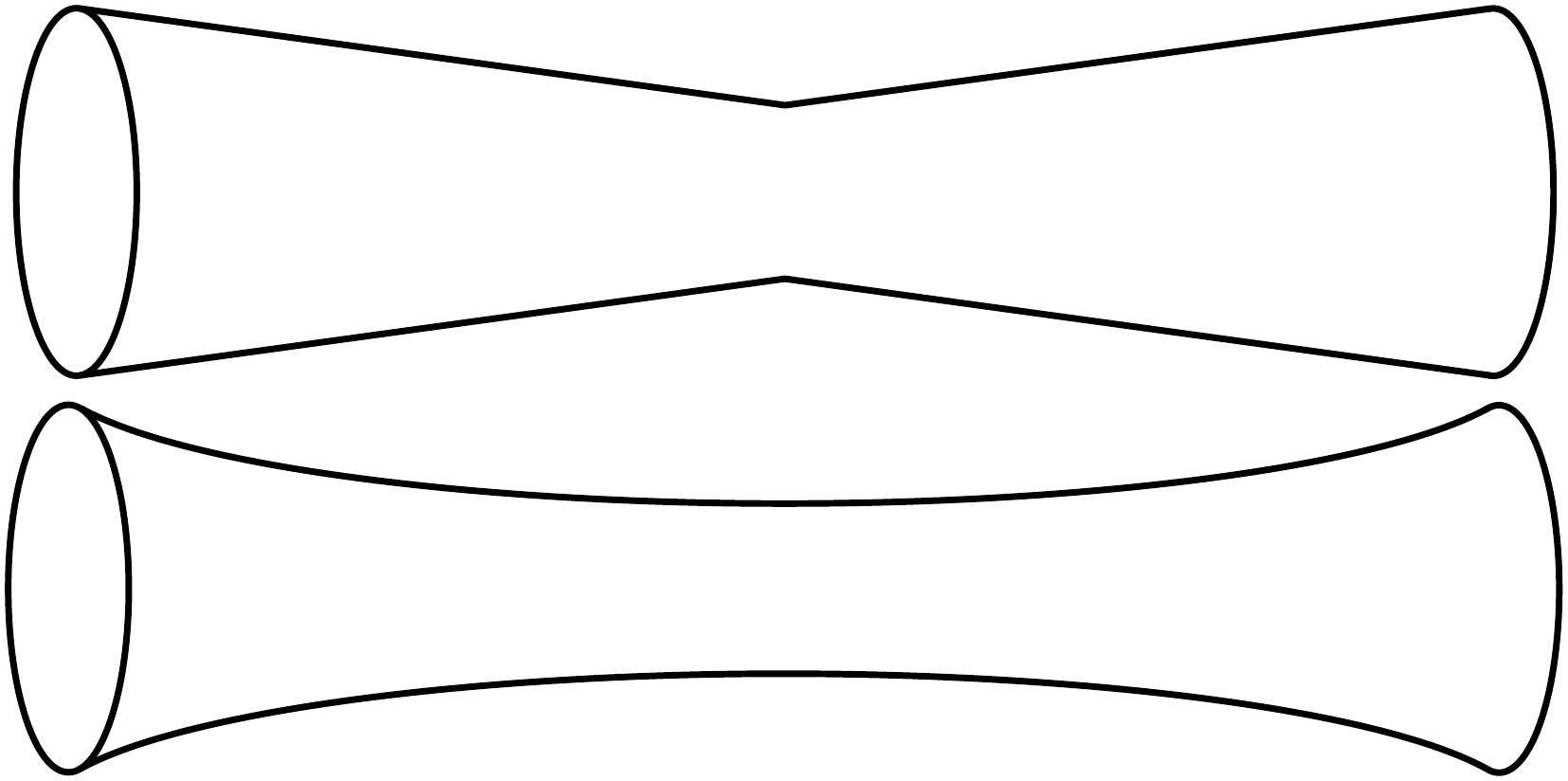} \caption{Profiles of converging-diverging axisymmetric capillaries.}
\label{Tubes}
\end{figure}

In this article we use the one-dimensional Navier-Stokes equations, which are widely used to
describe axisymmetric flows in large vessels, to derive analytical expressions for the flow of
Newtonian fluids in tubes of five axisymmetric converging-diverging geometries, some of which are
schematically depicted in Figure \ref{Tubes}, and compare our results to previously-derived
expressions using the lubrication approximation \cite{SochiNewtLub2010}. Other validation tests
have also been presented.

The widely-used one-dimensional form of the Navier-Stokes equations to describe the flow in a tube
of length $L$ where its axis is aligned with the $x$ axis and its midpoint is at $x=0$ is given by
the following continuity and momentum balance relations respectively assuming negligible
gravitational body forces \cite{FormaggiaGNQ2001, RuanCZC2003, SherwinFPP2003, UrquizaBLVF2003,
MilisicQ2004, FernandezMQ2005, FormaggiaLTV2006, FormaggiaMN2006, AlastrueyMPDPS2007,
AlastrueyPPS2008, LeeS2008, PasseriniLFQV2009, JanelaMS2010}

\begin{eqnarray}
\frac{\partial A}{\partial t}+\frac{\partial Q}{\partial x} & = & 0\,\,\,\,\,\,\,\,\,\,\,\,\, t\ge0,\,\,\, x\in\left[-\frac{L}{2},\frac{L}{2}\right]\\
\frac{\partial Q}{\partial t}+\frac{\partial}{\partial x}\left(\frac{\alpha
Q^{2}}{A}\right)+\frac{A}{\rho}\frac{\partial p}{\partial x}+\kappa\frac{Q}{A} & = &
0\,\,\,\,\,\,\,\,\,\,\,\,\, t\ge0,\,\,\, x\in\left[-\frac{L}{2},\frac{L}{2}\right]
\end{eqnarray}
In these equations, $A$ is the tube cross sectional area, $t$ is time, $Q$ ($=A\overline{u}$ with
$\overline{u}$ being the mean axial fluid speed) is the volumetric flow rate, $x$ is the axial
coordinate along the tube, $\alpha$ ($=\frac{\int u^{2}dA}{A\overline{u}^{2}}$ with $u$ being the
local axial fluid speed) \cite{FormaggiaLQ2003,SherwinFPP2003, FormaggiaLTV2006} is the correction
factor for axial momentum flux, $\rho$ is the fluid mass density, $p$ is the pressure, and $\kappa$
is a viscosity friction coefficient which is given by $\kappa=\frac{2\pi\alpha\mu}{\rho(\alpha-1)}$
\cite{AlastrueyMPDPS2007, LeeS2008} with $\mu$ being the fluid dynamic viscosity. This model is
considered one-dimensional due to the fact that the $\theta$ dependency of a
cylindrically-coordinated capillary is ignored due to the axisymmetric flow assumption while the
$r$ dependency is neglected because of the simplified consideration of the flow profile within a
lumped parameter which is the momentum correction factor. Therefore, the only explicitly-considered
dependency is the dependency in the flow direction, $x$.

For steady flow, the time terms are zero, and hence $Q$ as a function of $x$ is constant according
to the continuity equation. The momentum equation then becomes

\begin{equation}
\frac{\partial}{\partial x}\left(\frac{\alpha Q^{2}}{A}\right)+\frac{A}{\rho}\frac{\partial
p}{\partial x}+\kappa\frac{Q}{A}=0
\end{equation}
that is

\begin{equation}
\frac{\partial p}{\partial x}=-\frac{\rho}{A}\frac{\partial}{\partial x}\left(\frac{\alpha
Q^{2}}{A}\right)-\kappa\rho\frac{Q}{A^{2}}=\frac{\rho\alpha Q^{2}}{A^{3}}\frac{\partial A}{\partial
x}-\kappa\rho\frac{Q}{A^{2}}
\end{equation}

For a flow in the positive $x$ direction, the pressure gradient is negative and hence

\begin{equation}
p=\int_{X}\kappa\rho\frac{Q}{A^{2}}dx-\int_{X}\frac{\rho\alpha Q^{2}}{A^{3}}\frac{\partial
A}{\partial x}dx
\end{equation}

\begin{equation}
=\int_{X}\kappa\rho\frac{Q}{A^{2}}dx-\int_{A}\frac{\rho\alpha Q^{2}}{A^{3}}dA
\end{equation}

\begin{equation}
=\kappa\rho Q\int_{X}\frac{dx}{A^{2}}-\rho\alpha Q^{2}\int_{A}\frac{dA}{A^{3}}
\end{equation}
that is

\begin{equation}
p=\kappa\rho Q\int_{x=-L/2}^{L/2}\frac{dx}{A^{2}}+\frac{\rho\alpha
Q^{2}}{2}\left[\frac{1}{A^{2}}\right]_{x=-L/2}^{L/2}
\end{equation}

Due to the tube symmetry with respect to $x=0$

\begin{equation}
\int_{x=-L/2}^{L/2}\frac{dx}{A^{2}}=2\int_{x=0}^{L/2}\frac{dx}{A^{2}}
\end{equation}
and

\begin{equation}
\left[\frac{1}{A^{2}}\right]_{x=-L/2}^{L/2}=0\end{equation}

Hence

\begin{equation}
p=2\kappa\rho Q\int_{x=0}^{L/2}\frac{dx}{A^{2}}\label{PQEq}
\end{equation}

This expression is dimensionally consistent.

\subsection{Conical Tube} \label{}

\begin{figure}[!h]
\centering{}
\includegraphics
[scale=0.8] {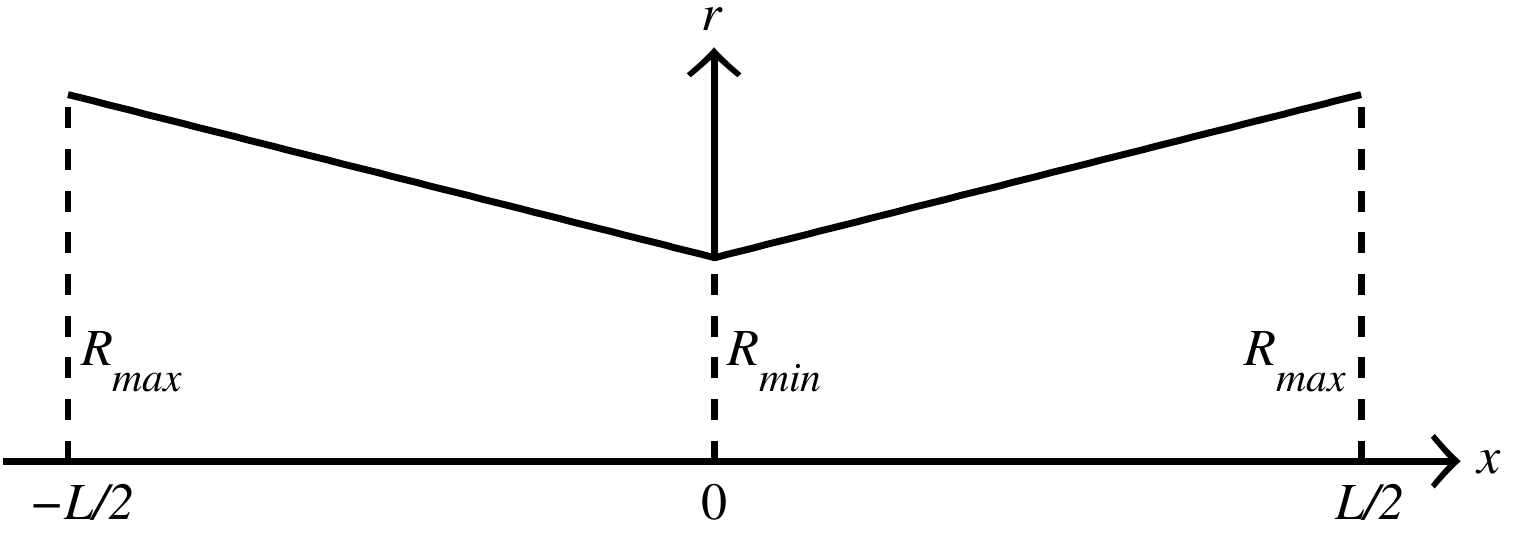} \caption{Schematic representation of the radius of a conically shaped
converging-diverging capillary as a function of the distance along the tube axis.} \label{Conic}
\end{figure}

For a tube of conical profile, depicted in Figure \ref{Conic}, the radius $r$ as a function of the
axial distance $x$ is given by

\begin{equation}
r(x)=a+b|x|\,\,\,\,\,\,\,\,-L/2\le x\le L/2
\end{equation}
where

\begin{equation}
a=R_{min}\hspace{1cm}{\rm and}\hspace{1cm}b=\frac{2(R_{max}-R_{min})}{L}
\end{equation}

Hence, Equation \ref{PQEq} becomes

\begin{equation} p=2\kappa\rho
Q\int_{x=0}^{L/2}\frac{dx}{A^{2}}=2\kappa\rho
Q\int_{x=0}^{L/2}\frac{dx}{\pi^{2}\left(a+bx\right)^{4}}
\end{equation}

\begin{equation}
=-2\kappa\rho Q\left[\frac{1}{3\pi^{2}b\left(a+bx\right)^{3}}\right]_{0}^{L/2}
\end{equation}

\begin{equation}
=-2\kappa\rho
Q\left[\frac{1}{3\pi^{2}\frac{2(R_{max}-R_{min})}{L}\left(R_{min}+\frac{2(R_{max}-R_{min})}{L}x\right)^{3}}\right]_{0}^{L/2}
\end{equation}

\begin{equation}
=-2\kappa\rho
Q\left[\frac{L}{6\pi^{2}(R_{max}-R_{min})R_{max}^{3}}-\frac{L}{6\pi^{2}(R_{max}-R_{min})R_{min}^{3}}\right]
\end{equation}
that is

\begin{equation}\label{fConic}
p=\frac{\kappa\rho
QL}{3\pi^{2}(R_{max}-R_{min})}\left[\frac{1}{R_{min}^{3}}-\frac{1}{R_{max}^{3}}\right]
\end{equation}

\subsection{Parabolic Tube} \label{}

\begin{figure}[!h]
\centering{}
\includegraphics
[scale=0.8] {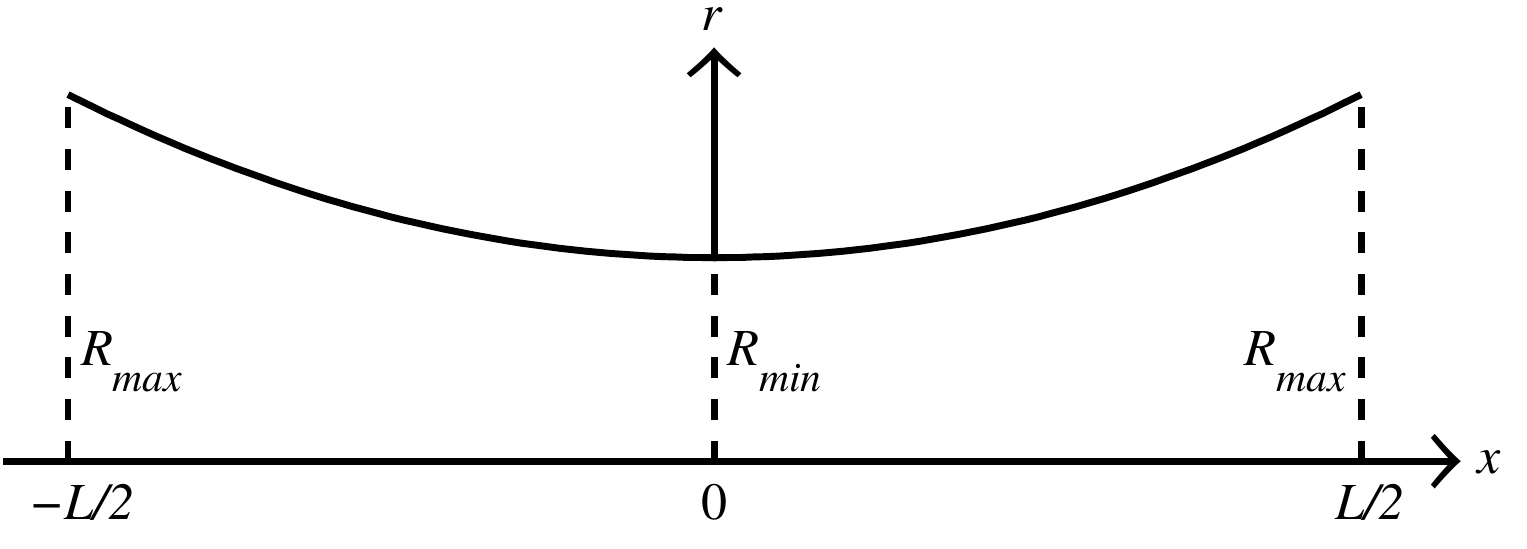} \caption{Schematic representation of the radius of a converging-diverging
capillary with a parabolic profile as a function of the distance along the tube axis.}
\label{Parabola}
\end{figure}

For a tube of parabolic profile, depicted in Figure \ref{Parabola}, the radius is given by

\begin{equation}
r(x)=a+bx^{2}\hspace{1cm}-L/2\le x\le L/2
\end{equation}
where

\begin{equation}
a=R_{min}\hspace{1cm}{\rm
and}\hspace{1cm}b=\left(\frac{2}{L}\right)^{2}(R_{max}-R_{min})
\end{equation}

Therefore, Equation \ref{PQEq} becomes

\begin{equation}
p=2\kappa\rho Q\int_{x=0}^{L/2}\frac{dx}{A^{2}}=2\kappa\rho
Q\int_{x=0}^{L/2}\frac{dx}{\pi^{2}\left(a+bx^{2}\right)^{4}}
\end{equation}

\begin{equation}
=\frac{2\kappa\rho
Q}{\pi^{2}}\left[\frac{x}{6a\left(a+bx^{2}\right)^{3}}+\frac{5x}{24a^{2}\left(a+bx^{2}\right)^{2}}+\frac{5x}{16a^{3}\left(a+bx^{2}\right)}+\frac{5\arctan\left(x\sqrt{\frac{b}{a}}\right)}{16a^{7/2}\sqrt{b}}\right]_{0}^{L/2}
\end{equation}
that is

\begin{equation}\label{fParabolic}
p=\frac{\kappa\rho
QL}{2\pi^{2}}\left[\frac{1}{3R_{min}R_{max}^{3}}+\frac{5}{12R_{min}^{2}R_{max}^{2}}+\frac{5}{8R_{min}^{3}R_{max}}+\frac{5\arctan\left(\sqrt{\frac{R_{max}-R_{min}}{R_{min}}}\right)}{8R_{min}^{7/2}\sqrt{R_{max}-R_{min}}}\right]
\end{equation}

\subsection{Hyperbolic Tube} \label{}

For a tube of hyperbolic profile, similar to the profile in Figure \ref{Parabola}, the radius is
given by

\begin{equation}
r(x)=\sqrt{a+bx^{2}}\hspace{1cm}-L/2\le x\le L/2\hspace{1cm}a,b>0
\end{equation}
where

\begin{equation}
a=R_{min}^{2}\hspace{1cm}{\rm
and}\hspace{1cm}b=\left(\frac{2}{L}\right)^{2}(R_{max}^{2}-R_{min}^{2})
\end{equation}

Therefore, Equation \ref{PQEq} becomes

\begin{equation}
p=2\kappa\rho Q\int_{x=0}^{L/2}\frac{dx}{A^{2}}=2\kappa\rho
Q\int_{x=0}^{L/2}\frac{dx}{\pi^{2}\left(a+bx^{2}\right)^{2}}
\end{equation}

\begin{equation}
=\frac{2\kappa\rho
Q}{\pi^{2}}\left[\frac{x}{2a(a+bx^{2})}+\frac{\arctan(x\sqrt{b/a})}{2a\sqrt{ab}}\right]_{0}^{L/2}
\end{equation}
that is

\begin{equation}\label{fHyperbolic}
p=\frac{\kappa\rho
QL}{2\pi^{2}}\left[\frac{1}{R_{min}^{2}R_{max}^{2}}+\frac{\arctan\left(\sqrt{\frac{R_{max}^{2}-R_{min}^{2}}{R_{min}^{2}}}\right)}{R_{min}^{3}\sqrt{R_{max}^{2}-R_{min}^{2}}}\right]
\end{equation}

\subsection{Hyperbolic Cosine Tube} \label{}

For a tube of hyperbolic cosine profile, similar to the profile in Figure \ref{Parabola}, the
radius is given by

\begin{equation}
r(x)=a\cosh(bx)\hspace{1cm}-L/2\le x\le L/2
\end{equation}
where

\begin{equation}
a=R_{min}\hspace{1cm}{\rm
and}\hspace{1cm}b=\frac{2}{L}\\\arccosh\left(\frac{R_{max}}{R_{min}}\right)
\end{equation}

Hence, Equation \ref{PQEq} becomes

\begin{equation}
p=2\kappa\rho Q\int_{x=0}^{L/2}\frac{dx}{A^{2}}=2\kappa\rho
Q\int_{x=0}^{L/2}\frac{dx}{\pi^{2}a^{4}\cosh^{4}(bx)}
\end{equation}

\begin{equation}
=\frac{2\kappa\rho
Q}{\pi^{2}}\left[\frac{\tanh(bx)\left[\mathrm{sech}^{2}(bx)+2\right]}{3a^{4}b}\right]_{0}^{L/2}
\end{equation}
that is

\begin{equation}\label{fCoshine}
p=\frac{\kappa\rho
QL}{3\pi^{2}}\left[\frac{\tanh\left(\\\arccosh\left(\frac{R_{max}}{R_{min}}\right)\right)\left[\mathrm{sech}^{2}\left(\\\arccosh\left(\frac{R_{max}}{R_{min}}\right)\right)+2\right]}{R_{min}^{4}\\\arccosh\left(\frac{R_{max}}{R_{min}}\right)}\right]
\end{equation}

\subsection{Sinusoidal Tube} \label{}

\begin{figure}[!h]
\centering{}
\includegraphics
[scale=0.8] {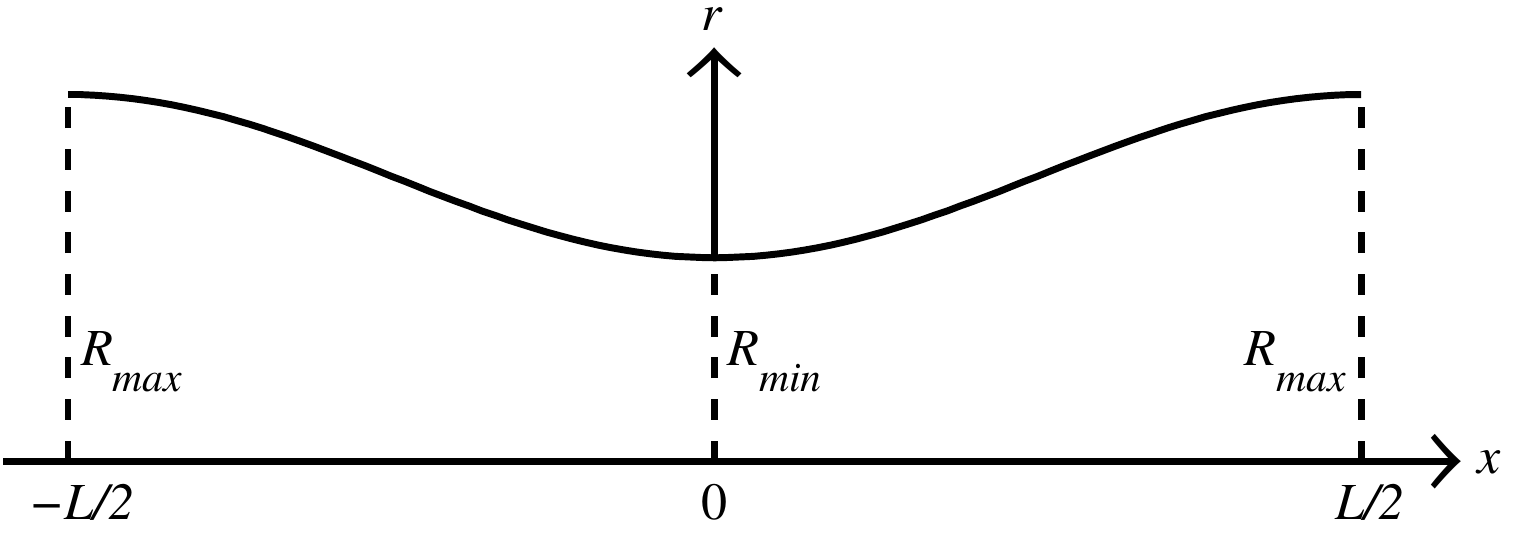} \caption{Schematic representation of the radius of a converging-diverging
capillary with a sinusoidal profile as a function of the distance along the tube axis.}
\label{Sinusoid}
\end{figure}

For a tube of sinusoidal profile, depicted in Figure \ref{Sinusoid}, where the tube length $L$
spans one complete wavelength, the radius is given by

\begin{equation}
r(x)=a-b\cos\left(kx\right)\hspace{1cm}-L/2\le x\le L/2\hspace{1cm}a>b>0
\end{equation}
where

\begin{equation}
a=\frac{R_{max}+R_{min}}{2}\hspace{1cm}b=\frac{R_{max}-R_{min}}{2}\hspace{1cm}\&\hspace{1cm}k=\frac{2\pi}{L}
\end{equation}

Hence, Equation \ref{PQEq} becomes

\begin{equation}
p=2\kappa\rho Q\int_{x=0}^{L/2}\frac{dx}{A^{2}}=2\kappa\rho
Q\int_{x=0}^{L/2}\frac{dx}{\pi^{2}\left[a-b\cos\left(kx\right)\right]^{4}}
\end{equation}

On performing this integration, the following relation is obtained

\begin{equation}
p=\frac{2\kappa\rho Q}{\pi^{2}b^{4}k}\left[I\right]_{0}^{L/2}
\end{equation}
where

\begin{eqnarray}
\nonumber I & = &
\frac{(6B^{3}+9B)}{3(B^{2}-1)^{7/2}}\arctan\left(\frac{(B-1)\tan(\frac{kx}{2})}{\sqrt{B^{2}-1}}\right)-\frac{(11B^{2}+4)\sin(kx)}{6(B^{2}-1)^{3}[B+\cos(kx)]}\\
 &  & -\frac{5B\sin(kx)}{6(B^{2}-1)^{2}[B+\cos(kx)]^{2}}-\frac{\sin(kx)}{3(B^{2}-1)[B+\cos(kx)]^{3}}
\end{eqnarray}

\begin{equation}
\&\hspace{1cm}B=\frac{R_{max}+R_{min}}{R_{min}-R_{max}}
\end{equation}

On taking $\lim_{x\rightarrow\frac{L}{2}^{-}}I$ the following expression is obtained

\begin{equation}
p=\frac{2\kappa\rho
Q}{\pi^{2}b^{4}k}\left[-\frac{(6B^{3}+9B)}{3(B^{2}-1)^{7/2}}\frac{\pi}{2}\right]=-\frac{\kappa\rho
Q(6B^{3}+9B)}{3\pi b^{4}k(B^{2}-1)^{7/2}}
\end{equation}

Since $B<-1$, $p>0$ as it should be. On substituting for $B$, $b$ and $k$ in the last expression we
obtain

\begin{equation}\label{fSinusoid}
p=\frac{\kappa\rho
QL\left(R_{max}-R_{min}\right)^{3}\left[2\left(\frac{R_{max}+R_{min}}{R_{max}-R_{min}}\right)^{3}+3\left(\frac{R_{max}+R_{min}}{R_{max}-R_{min}}\right)\right]}{16\pi^{2}\left(R_{max}R_{min}\right)^{7/2}}
\end{equation}

\vspace{0.5cm}

It is noteworthy that all these relations (i.e. Equations \ref{fConic}, \ref{fParabolic},
\ref{fHyperbolic}, \ref{fCoshine} and \ref{fSinusoid}), are dimensionally consistent.

\section{Validation} \label{Validation}

The derived converging-diverging formulae (Equations \ref{fConic}, \ref{fParabolic},
\ref{fHyperbolic}, \ref{fCoshine} and \ref{fSinusoid}) have been validated by three different ways:
comparison to previously-derived formulae based on lubrication approximation, numerical
integration, and testing the convergence behavior of the analytical solutions in the limiting cases
where the converging-diverging geometries approach a straight tube geometry. These validation
methods are outlined in the following subsections.

\subsection{Comparison to Lubrication Formulae}

For $\alpha=4/3$, the derived five relations are identical to the previously-derived expressions
using the lubrication approximation (refer to Table \ref{lubTable}). Since $\alpha=4/3$ corresponds
to a fully-developed parabolic velocity profile (Poiseuille flow), the lubrication approximation
formulae are special cases of the more general 1D Navier-Stokes relations which can accommodate
other velocity profiles, such as flat profile (plug flow) when $\alpha$ approaches unity
\cite{FormaggiaGNQ2001, SmithPH2002, SherwinFPP2003, FormaggiaMN2006, FormaggiaLTV2006, LeeS2008}.

\vspace{0.5cm}

\begin{table} [!h]
\centering %
\caption[Lubrication approximation table]{Lubrication approximation table. These formulae are
derived in \cite{SochiNewtLub2010}.}
\label{lubTable} %
\vspace{0.5cm} %
{\scriptsize
\begin{tabular}{|l|l|}
\hline Conical &
$p=\frac{8LQ\mu}{3\pi(R_{max}-R_{min})}\left(\frac{1}{R_{min}^{3}}-\frac{1}{R_{max}^{3}}\right)$\tabularnewline
\hline Parabolic &
$p=\frac{4LQ\mu}{\pi}\left(\frac{1}{3R_{min}R_{max}^{3}}+\frac{5}{12R_{min}^{2}R_{max}^{2}}+\frac{5}{8R_{min}^{3}R_{max}}+\frac{5\arctan\left(\sqrt{\frac{R_{max}-R_{min}}{R_{min}}}\right)}{8R_{min}^{7/2}\sqrt{R_{max}-R_{min}}}\right)$\tabularnewline
\hline Hyperbolic &
$p=\frac{4LQ\mu}{\pi}\left(\frac{1}{R_{min}^{2}R_{max}^{2}}+\frac{\arctan\left(\sqrt{\frac{R_{max}^{2}-R_{min}^{2}}{R_{min}^{2}}}\right)}{R_{min}^{3}\sqrt{R_{max}^{2}-R_{min}^{2}}}\right)$\tabularnewline
\hline Hyperbolic Cosine & $p=\frac{8LQ\mu}{3\pi
R_{min}^{4}}\left(\frac{\tanh\left(\arccosh\left(\frac{R_{max}}{R_{min}}\right)\right)\left\{
\mathrm{sech}^{2}\left(\arccosh\left(\frac{R_{max}}{R_{min}}\right)\right)+2\right\}
}{\arccosh\left(\frac{R_{max}}{R_{min}}\right)}\right)$\tabularnewline \hline Sinusoidal &
$p=\frac{LQ\mu\left\{ 2(R_{max}+R_{min})^{3}+3(R_{max}+R_{min})(R_{max}-R_{min})^{2}\right\}
}{2\pi(R_{max}R_{min})^{7/2}}$\tabularnewline \hline
\end{tabular}
}
\end{table}

\subsection{Numerical Integration}

\begin{figure}
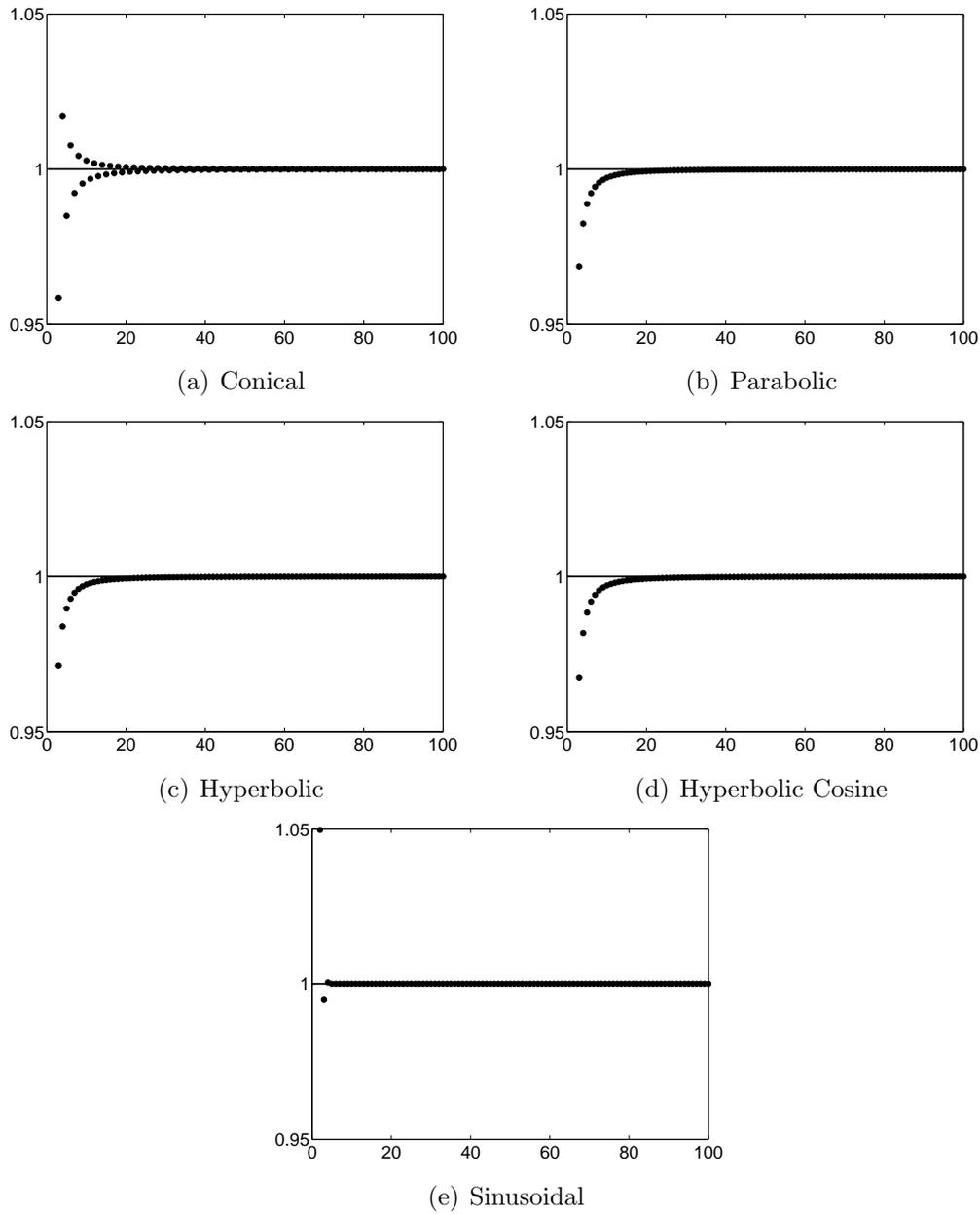

\centering %
\subfigure[Conical]%
{\begin{minipage}[b]{0.5\textwidth} \CIF {VConic}
\end{minipage}}
\Hs %
\subfigure[Parabolic]%
{\begin{minipage}[b]{0.5\textwidth} \CIF {VParabola}
\end{minipage}} \Vmin

%
\centering %
\subfigure[Hyperbolic]%
{\begin{minipage}[b]{0.5\textwidth} \CIF {VHyperbola}
\end{minipage}}
\Hs %
\subfigure[Hyperbolic Cosine]%
{\begin{minipage}[b]{0.5\textwidth} \CIF {VHyperbolicCosine}
\end{minipage}} \Vmin

%
\centering %
\subfigure[Sinusoidal]%
{\begin{minipage}[b]{0.5\textwidth} \CIF {VSinusoidal}
\end{minipage}}
\caption{The ratio of numeric to analytic solutions ($y$-axis) as a function of the number of
discretization elements ($x$-axis) for the five converging-diverging geometries using typical
values for the flow, fluid and capillary parameters. Similar convergence behavior has been observed
for other sets of parameters. \label{ValidationPlots}}
\end{figure}

The derived converging-diverging formulae have also been validated by comparison to numerical
solutions based on numerical integration by discretizing the converging-diverging tubes of these
geometries and solving for a Newtonian flow in each element by averaging the radius of that
element. This method has been applied to these geometries using diverse sets of fluid, flow and
tube parameters which include $\mu$, $\rho$, $p$, $Q$, $\alpha$, $L$, $R_{min}$, and $R_{max}$. A
sample of these numerical validations is presented in Figure~\ref{ValidationPlots} for the five
geometries using typical parameters. As seen, the numerical solutions converge to the analytical
solutions fairly quickly in all cases. The two solutions become virtually identical for a typical
meshing of 40--50 elements. The qualitative difference in convergence behavior between the conical
and sinusoidal on one hand and the other geometries on the other hand seems to arise from the
converging-diverging nature of these geometries and how gradually it takes place over the tube
length. These two types of observed convergence behavior (i.e. oscillatory like conical and
asymptotic like parabolic) occur in various numerical contexts and have been observed in different
numerical implementations by the author and by other researchers. Another remark is that the
convergence rate indicates the quality of the average radius as an indicator of the effective
radius of the element. The rapid convergence of the sinusoidal tube may support this guess as the
smoothness of the sinusoidal profile makes the average radius very good representative of the
effective radius of the discretized sinusoidal segments.

\subsection{Convergence to Straight Tube Solution}

For $\alpha=4/3$ (parabolic velocity profile) the sinusoidal equation reduces to the Poiseuille
expression when $R_{max}=R_{min}$. With regard to the other geometries, despite the fact that the
other four equations are not defined when $R_{max}=R_{min}$, they converge to the Poiseuille value
as $R_{max}$ approaches $R_{min}$, and hence for all practical purposes they are numerically
identical to the Poiseuille value when the difference between the two radii is negligible.

\newpage
\section{Conclusions} \label{Conclusions}

In this paper we derived analytical expressions relating the pressure drop to the volumetric flow
rate for Newtonian fluids in five different converging-diverging geometries using the
one-dimensional Navier-Stokes flow equations in axisymmetric rigid tubes. The results obtained in
this paper are identical, within a non-dimensional numerical factor, to those derived in
\cite{SochiNewtLub2010} using the lubrication approximation. The results have also been validated
numerically by comparing the analytical solutions to numerical results obtained from numerical
integration for a wide range of flow, fluid and tube characteristics. Moreover, they have been
endorsed by testing the convergence behavior of the analytical solutions as the
converging-diverging geometries approach the limiting case of a straight tube geometry when
$R_{max} \rightarrow R_{min}$.

These expressions can be used in various practical scientific and engineering situations to
describe isothermal, uniform, laminar, time-independent flow of incompressible Newtonian fluids
through converging-diverging flow paths. These situations include the flow in corrugated vessels
and the flow in the pores and throats of porous media where the converging-diverging nature can be
idealized by these relatively-simple geometries. The analytical method can also be used to derive
expressions for geometries other than those presented in this paper.

\newpage
\phantomsection \addcontentsline{toc}{section}{Nomenclature} %
{\noindent \LARGE \bf Nomenclature} \vspace{0.5cm}

\begin{supertabular}{ll}
$\alpha$                &   correction factor for axial momentum flux \\
$\kappa$                &   viscosity friction coefficient (m$^2$.s$^{-1}$) \\
$\mu$                   &   fluid dynamic viscosity (Pa.s) \\
$\rho$                  &   fluid mass density (kg.m$^{-3}$) \\
\\
$A$                     &   tube cross sectional area (m$^{2}$) \\
$L$                     &   tube length (m) \\
$p$                     &   pressure (Pa) \\
$Q$                     &   volumetric flow rate (m$^{3}$.s$^{-1}$) \\
$r$                     &   tube radius (m) \\
$R_{max}$               &   maximum radius of converging-diverging tube (m) \\
$R_{min}$               &   minimum radius of converging-diverging  tube (m) \\
$t$                     &   time (s) \\
$u$                     &   local axial fluid speed (m.s$^{-1}$) \\
$\overline{u}$          &   mean axial fluid speed (m.s$^{-1}$) \\
$x$                     &   axial coordinate (m) \\
\end{supertabular}

\newpage
\phantomsection \addcontentsline{toc}{section}{References} %
\bibliographystyle{unsrt}

\newpage

\begin{thebibliography}{10}

\bibitem{SochiB2008}
T.~Sochi;~M.J. Blunt.
\newblock {Pore-scale network modeling of Ellis and Herschel-Bulkley fluids}.
\newblock {\em Journal of Petroleum Science and Engineering}, 60(2):105--124,
  2008.

\bibitem{SochiVE2009}
T.~Sochi.
\newblock {Pore-scale modeling of viscoelastic flow in porous media using a
  Bautista-Manero fluid}.
\newblock {\em International Journal of Heat and Fluid Flow}, 30(6):1202--1217,
  2009.

\bibitem{SochiFeature2010}
T.~Sochi.
\newblock {Non-Newtonian Flow in Porous Media}.
\newblock {\em Polymer}, 51(22):5007--5023, 2010.

\bibitem{SochiYield2010}
T.~Sochi.
\newblock {Modelling the Flow of Yield-Stress Fluids in Porous Media}.
\newblock {\em Transport in Porous Media}, 85(2):489--503, 2010.

\bibitem{SochiComp2010}
T.~Sochi.
\newblock {Computational Techniques for Modeling Non-Newtonian Flow in Porous
  Media}.
\newblock {\em International Journal of Modeling, Simulation, and Scientific
  Computing}, 1(2):239--256, 2010.

\bibitem{SochiArticle2010}
T.~Sochi.
\newblock {Flow of Non-Newtonian Fluids in Porous Media}.
\newblock {\em Journal of Polymer Science Part B}, 48(23):2437--2467, 2010.

\bibitem{SochiPower2011}
T.~Sochi.
\newblock {The flow of power-law fluids in axisymmetric corrugated tubes}.
\newblock {\em Journal of Petroleum Science and Engineering}, 78(3-4):582--585,
  2011.

\bibitem{SochiSlip2011}
T.~Sochi.
\newblock {Slip at Fluid-Solid Interface}.
\newblock {\em Polymer Reviews}, 51:1--33, 2011.

\bibitem{Whitebook1991}
F.M. White.
\newblock {\em {Viscous Fluid Flow}}.
\newblock McGraw Hill Inc., second edition, 1991.

\bibitem{SisavathJZ2001}
S.~Sisavath; X. Jing;~R.W. Zimmerman.
\newblock {Laminar Flow Through Irregularly-Shaped Pores in Sedimentary Rocks}.
\newblock {\em Transport in Porous Media}, 45(1):41--62, 2001.

\bibitem{KozickiCT1966}
W.~Kozicki; C.H. Chou;~C. Tiu.
\newblock {Non-Newtonian flow in ducts of arbitrary cross-sectional shape}.
\newblock {\em Chemical Engineering Science}, 21(8):665--679, 1966.

\bibitem{Miller1972}
C.~Miller.
\newblock {Predicting Non-Newtonian Flow Behavior in Ducts of Unusual Cross
  Section}.
\newblock {\em Industrial \& Engineering Chemistry Fundamentals},
  11(4):524--528, 1972.

\bibitem{Oka1973}
S.~Oka.
\newblock {Pressure development in a non-Newtonian flow through a tapered
  tube}.
\newblock {\em Rheologica Acta}, 12(2):224--227, 1973.

\bibitem{WilliamsJ1980}
E.W. Williams;~S.H. Javadpour.
\newblock {The flow of an elastico-viscous liquid in an axisymmetric pipe of
  slowly varying cross-section}.
\newblock {\em Journal of Non-Newtonian Fluid Mechanics}, 7(2-3):171--188,
  1980.

\bibitem{ThienGB1985}
N.~Phan-Thien; C.J. Goh;~M.B. Bush.
\newblock {Viscous flow through corrugated tube by boundary element method}.
\newblock {\em Journal of Applied Mathematics and Physics (ZAMP)},
  36(3):475--480, 1985.

\bibitem{ThienK1987}
N.~Phan-Thien;~M.M.K. Khan.
\newblock {Flow of an Oldroyd-type fluid through a sinusoidally corrugated
  tube}.
\newblock {\em Journal of Non-Newtonian Fluid Mechanics}, 24(2):203--220, 1987.

\bibitem{LahbabiC1986}
A.~Lahbabi; H-C. Chang.
\newblock {Flow in periodically constricted tubes: Transition to inertial and
  nonsteady flows}.
\newblock {\em Chemical Engineering Science}, 41(10):2487--2505, 1986.

\bibitem{BurdetteCAB1989}
S.R. Burdette; P.J. Coates; R.C. Armstrong;~R.A. Brown.
\newblock {Calculations of viscoelastic flow through an axisymmetric corrugated
  tube using the explicitly elliptic momentum equation formulation (EEME)}.
\newblock {\em Journal of Non-Newtonian Fluid Mechanics}, 33(1):1--23, 1989.

\bibitem{PilitsisSB1991}
S.~Pilitsis; A. Souvaliotis;~A.N. Beris.
\newblock {Viscoelastic flow in a periodically constricted tube: The combined
  effect of inertia, shear thinning, and elasticity}.
\newblock {\em Journal of Rheology}, 35(4):605--646, 1991.

\bibitem{PilitsisB1989}
S.~Pilitsis;~A.N. Beris.
\newblock {Calculations of steady-state viscoelastic flow in an undulating
  tube}.
\newblock {\em Journal of Non-Newtonian Fluid Mechanics}, 31(3):231--287, 1989.

\bibitem{JamesTKBP1990}
D.F. James; N. Phan-Thien; M.M.K. Khan; A.N. Beris;~S. Pilitsis.
\newblock {Flow of test fluid M1 in corrugated tubes}.
\newblock {\em Journal of Non-Newtonian Fluid Mechanics}, 35(2-3):405--412,
  1990.

\bibitem{TalwarK1992}
K.K. Talwar;~B. Khomami.
\newblock {Application of higher order finite element methods to viscoelastic
  flow in porous media}.
\newblock {\em Journal of Rheology}, 36(7):1377--1416, 1992.

\bibitem{KoshibaMNS2000}
T.~Koshiba; N. Mori; K. Nakamura;~S. Sugiyama.
\newblock {Measurement of pressure loss and observation of the flow field in
  viscoelastic flow through an undulating channel}.
\newblock {\em Journal of Rheology}, 44(1):65--78, 2000.

\bibitem{MasulehP2004}
S.H. Momeni-Masuleh;~T.N. Phillips.
\newblock {Viscoelastic flow in an undulating tube using spectral methods}.
\newblock {\em Computers \& fluids}, 33(8):1075--1095, 2004.

\bibitem{DavidsonLC2008}
D.~Davidson; G.L. Lehmann;~E.J. Cotts.
\newblock {Horizontal capillary flow of a Newtonian liquid in a narrow gap
  between a plane wall and a sinusoidal wall}.
\newblock {\em Fluid Dynamics Research}, 40(11-12):779--802, 2008.

\bibitem{SochiNewtLub2010}
T.~Sochi.
\newblock {The Flow of Newtonian Fluids in Axisymmetric Corrugated Tubes}.
\newblock 2010.
\newblock arXiv:1006.1515v1.

\bibitem{FormaggiaGNQ2001}
L.~Formaggia; J.F. Gerbeau; F. Nobile;~A. Quarteroni.
\newblock {On the coupling of 3D and 1D Navier-Stokes equations for flow
  problems in compliant vessels}.
\newblock {\em Computer Methods in Applied Mechanics and Engineering},
  191(6-7):561--582, 2001.

\bibitem{RuanCZC2003}
W.~Ruan; M.E. Clark; M. Zhao;~A. Curcio.
\newblock {A Hyperbolic System of Equations of Blood Flow in an Arterial
  Network}.
\newblock {\em SIAM Journal on Applied Mathematics}, 64(2):637--667, 2003.

\bibitem{SherwinFPP2003}
S.J. Sherwin; V. Franke; J. Peir\'{o};~K. Parker.
\newblock {One-dimensional modelling of a vascular network in space-time
  variables}.
\newblock {\em Journal of Engineering Mathematics}, 47(3-4):217--250, 2003.

\bibitem{UrquizaBLVF2003}
S.~Urquiza; P. Blanco; G. Lombera; M. Venere;~R. Feijoo.
\newblock {Coupling Multidimensional Compliant Models For Carotid Artery Blood
  Flow}.
\newblock {\em Mec\'{a}nica Computacional}, XXII(3):232--243, 2003.

\bibitem{MilisicQ2004}
V.~Mili\v{s}i\'{c};~A. Quarteroni.
\newblock {Analysis of lumped parameter models for blood flow simulations and
  their relation with 1D models}.
\newblock {\em Mathematical Modelling and Numerical Analysis}, 38(4):613--632,
  2004.

\bibitem{FernandezMQ2005}
M.\'{A}. Fern\'{a}ndez; V. Mili\v{s}i\'{c};~A. Quarteroni.
\newblock {Analysis of a Geometrical Multiscale Blood Flow Model Based on the
  Coupling of ODEs and Hyperbolic PDEs}.
\newblock {\em Multiscale Modeling \& Simulation}, 4(1):215--236, 2005.

\bibitem{FormaggiaLTV2006}
L.~Formaggia; D. Lamponi; M. Tuveri;~A. Veneziani.
\newblock {Numerical modeling of 1D arterial networks coupled with a lumped
  parameters description of the heart}.
\newblock {\em Computer Methods in Biomechanics and Biomedical Engineering},
  9(5):273--288, 2006.

\bibitem{FormaggiaMN2006}
L.~Formaggia; A. Moura;~F. Nobile.
\newblock {Coupling 3D and 1D fluid-structure interaction models for blood flow
  simulations}.
\newblock {\em Proceedings in Applied Mathematics and Mechanics, Special Issue:
  GAMM Annual Meeting 2006 - Berlin}, 6(1):27--30, 2006.

\bibitem{AlastrueyMPDPS2007}
J.~Alastruey; S.M. Moore; K.H. Parker; T. David; J. Peir\'{o}~S.J. Sherwin.
\newblock {Reduced modelling of blood flow in the cerebral circulation:
  Coupling 1-D, 0-D and cerebral auto-regulation models}.
\newblock {\em International Journal for Numerical Methods in Fluids},
  56(8):1061--1067, 2008.

\bibitem{AlastrueyPPS2008}
J.~Alastruey; K.H. Parker; J. Peir\'{o};~S.J. Sherwin.
\newblock {Lumped parameter outflow models for 1-D blood flow simulations:
  Effect on pulse waves and parameter estimation}.
\newblock {\em Communications in Computational Physics}, 4(2):317--336, 2008.

\bibitem{LeeS2008}
J.~Lee;~N. Smith.
\newblock {Development and application of a one-dimensional blood flow model
  for microvascular networks}.
\newblock {\em Proceedings of the Institution of Mechanical Engineers, Part H:
  Journal of Engineering in Medicine}, 222(4):487--512, 2008.

\bibitem{PasseriniLFQV2009}
T.~Passerini;~M. de~Luca; L. Formaggia; A. Quarteroni; A.~Veneziani.
\newblock {A 3D/1D geometrical multiscale model of cerebral vasculature}.
\newblock {\em Journal of Engineering Mathematics}, 64(4):319--330, 2009.

\bibitem{JanelaMS2010}
J.~Janela;~A.B. de~Moura; A.~Sequeira.
\newblock {Comparing Absorbing Boundary Conditions for a 3D Non Newtonian
  Fluid-Structure Interaction Model for Blood Flow in Arteries}.
\newblock {\em Mec\'{a}nica Computacional}, XXIX(59):5961--5971, 2010.

\bibitem{FormaggiaLQ2003}
L.~Formaggia; D. Lamponi;~A. Quarteroni.
\newblock {One-dimensional models for blood flow in arteries}.
\newblock {\em Journal of Engineering Mathematics}, 47(3/4):251--276, 2003.

\bibitem{SmithPH2002}
N.P. Smith; A.J. Pullan;~P.J. Hunter.
\newblock {An Anatomically Based Model of Transient Coronary Blood Flow in the
  Heart}.
\newblock {\em SIAM Journal on Applied Mathematics}, 62(3):990--1018, 2002.

\end{thebibliography}

\end{document}